%% file: wiki_evolution_arxiv.tex
\documentclass{sig-alternate}

\usepackage{graphicx}
\usepackage{url}
\usepackage[utf8]{inputenc}
\usepackage{psfrag}
\usepackage{subfigure}

\newenvironment{packed_itemize}{
\begin{itemize}
 \setlength{\itemsep}{1pt}
 \setlength{\parskip}{0pt}
 \setlength{\parsep}{0pt}
}{\end{itemize}}

\newenvironment{packed_enumerate}{
\begin{enumerate}
 \setlength{\itemsep}{1pt}
 \setlength{\parskip}{0pt}
 \setlength{\parsep}{0pt}
}{\end{enumerate}}

\newcommand{\mpara}[1]{\medskip\noindent{\bf #1}}

\def\sharedaffiliation{%
\end{tabular}
\begin{tabular}{c}}

\permission{}
\copyrightetc{}

\begin{document}

\title{There is No Deadline - Time Evolution of Wikipedia Discussions}
\numberofauthors{3} \author{
  \alignauthor Andreas Kaltenbrunner\\
  \email{\normalsize{kaltenbrunner@gmail.com}} \alignauthor
  \alignauthor David Laniado\\
  \email{\normalsize{david.laniado@barcelonamedia.org}}
  \sharedaffiliation
  \affaddr{Barcelona Media Foundation} \\
  \affaddr{Barcelona, Spain} }

\maketitle
\begin{abstract}
  Wikipedia articles are by definition never finished: at any moment
  their content can be edited, or discussed in the associated talk
  pages.  In this study we analyse the evolution of these discussions
  to unveil patterns of collective participation along the temporal
  dimension, and to shed light on the process of content creation on
  different topics.  At a micro-scale, we investigate peaks in the
  discussion activity and we observe a non-trivial relationship with
  edit activity.  At a larger scale, we introduce a measure to account
  for how fast discussions grow in complexity, and we find speeds that
  span three orders of magnitude for different articles.  Our analysis
  should help the community in tasks such as early detection of
  controversies and assessment of discussion maturity.
\end{abstract}
\category{H.5.3}{Information Interfaces}{Group and
  Organization Interfaces}[Computer-supported cooperative work,
Web-based interaction]
\keywords{Wikipedia, online discussion, temporal patterns, h-index} 

\section{Introduction}
Everyday thousands of people from all over the world edit the articles
of Wikipedia, and millions access their content to get information and
form opinions about the most various kinds of topics.  While one of
the weaknesses of traditional encyclopedias is the lack of up-to-date
information, due to the time required by the publishing process,
Wikipedia tends to cover also very recent events and fresh
information, as it can be updated in real time. This is one of its
most essential characteristics, as suggested by the very same word
\emph{wiki}, which in Hawaiian means \emph{quick}.

Yet, not all articles evolve with the same speed and whether in some
cases \emph{quick} may be \emph{too fast} is object of discussion
among the editors of Wikipedia. An essay invites users to consider
whether an addition will still appear relevant in ten
years\footnote{\url{http://en.wikipedia.org/wiki/Wikipedia:Recentism}}
to avoid \emph{recentism}, i.e. editing an article without a
long-term, historical view. Nevertheless, although another sibling
project has been created expressly for in-depth coverage of
news\footnote{\url{http://www.wikinews.org/}}, also for recent and
current events most editors seem to prefer Wikipedia articles, as
unique collective artifacts in constant evolution.

If one would like to study the speed with which an article changes,
the first natural choice would be to study the number of edits per
time unit. However, the larger an article becomes, more and more of
its generative process takes place in talk pages, parallel spaces to
every article, where editors are engaged in discussions with the
objective of reaching consensus on the article content. These
discussions can be seen as a magmatic flow of arguments and
counter-arguments, which are progressively transformed into the
encyclopedic content.  Looking at the associated discussion is often
the most effective way to understand the process which brought to the
current state of an article, showing traces of the explicit
coordination among editors in the form of suggestions, disputes,
complains, invocation of community policies, polls and pursuit of
consensus~\cite{Viegas2007talk}.

This motivates us to investigate the temporal patterns of these
discussions and their relation with edit activity. While previous
studies have focused on the amount of activity on talk
pages~\cite{kittur2007he,yasseri2012dynamics}, and the shape and
complexity of the discussions~\cite{Laniado2011}, still little is
known about the dynamics of these conversations.

How are the conversations distributed in time?  How frequent are
spikes of activity?  Are they associated to external events or to
community dynamics?  What is the relationship between discussion and
edit activity?  How fast do discussions grow, and for how long?  Does
the topic affect the way a discussion evolves?

Being able to answer these questions could benefit the community in
tasks like assessment of discussion maturity and early detection of
controversies.  Understanding when the biggest controversies
associated to an article have been solved (i.e. the discussion is
mature) may serve as a proxy for content stability, while detecting
disputes which may need the attention of moderators could help avoid
their escalation into a too fierce debate, allowing Wikipedians to
invest their time into more productive tasks and to reduce their
frustration, and ultimately might help to reduce editor drop-out.

To address the above-mentioned research questions, we present here an
extensive study of the temporal dynamics of the English Wikipedia
discussions, taking into account all the conversations on article talk
pages occurring during its first 9 years of history. We start by
summarising our main findings and offering a brief insight into
related work in sections~\ref{sec:contribution}
and~\ref{sec:related}. Then, after introducing the dataset in
Section~\ref{sec:data}, we detect and contrast peaks in the edit and
discussion activity of articles in Section~\ref{sec:peaks}.
Afterwards, in Section~\ref{sec:speed}, we introduce a metric based on
the h-index~\cite{hirsch05} to measure the growth speed of
discussions. As an example, we illustrate the evolution of the
articles about the last three US presidents, and apply our measure to
all heavily discussed articles to reveal differences in the
discussions about different subjects. 
Finally, in the last section we draw conclusions and
lines for future research.

\section{Our contribution}
\label{sec:contribution}
Our main contributions are:
\vspace{-2mm}
\begin{packed_itemize}
\item We detect peaks in the discussions on Wikipedia articles and
  shed light on some of their statistical properties such as number of peaks
  per article, peak length and time interval between peaks. A
  comparison with peaks in the edit history reveals that comment and
  edit peaks appear mainly independently of each other.
\item We offer a qualitative insight into the nature of edit and
  discussion peaks through the detailed analysis of a case study,
  which shows that both endogenous (Wikipedia internal) and exogenous
  (offline world) events can be the cause of such peaks, while the
  subjects of the disputes are not necessarily related to such events.
\item We introduce a measure to account for the speed with which the
  discussions gain in complexity. The measure reveals speeds that span
  three orders of magnitude.
\item We observe that articles about hot topics in the media receive
  also considerably more attention, have faster dynamics and longer
  peak durations in Wikipedia. %

\end{packed_itemize}

\section{Related work}
\label{sec:related}
Transparency is a pillar of the wiki philosophy, so complete
information about the edit history of each page is available. However,
it is often difficult to make sense of it, especially for larger
pages. To address this limitation, a number of tools have been
proposed for the visualisation of the evolution of the activity on a
given page~\cite{brandes2008visual,suh2008lifting,Viegas2007talk}.
Among the studies focused on mining edit history to extract patterns
of interaction, we mention~\cite{adler2011vandalism} for automatic
vandalism detection,~\cite{kittur2007he}
and~\cite{yasseri2012dynamics} for identifying conflict. The latter
also provides a description of the most controversial topics in
several language versions of Wikipedia; for the English Wikipedia,
positive correlation is found between conflict on an article and
length of the associated talk page.

While at the beginning most of the effort in Wikipedia was spent in
creating and editing articles, other activities such as coordination
and discussion have progressively emerged with the increased
complexity~\cite{kittur2007he,Suh09singularity}.  Kittur et
al.~\cite{kittur2007bourgeoisie} studied the distribution of work over
time, and found a gradual shift from the ``elite'' users to less
active editors, while the authors of~\cite{coauthorship2} described
the evolution of Wikipedia's co-authorship network pointing out its
increasing centralisation around the most active users.

The interpretation of Wikipedia as a global memory place led Ferron et
al.~\cite{ferron2000studying} to observe that articles related to
traumatic events are more likely to be characterised by a higher
amount of edits in correspondence with anniversaries.
In~\cite{ferron2011collective} the same authors present a study of
activity on different language versions of Wikipedia during the
Egyptian revolution, finding evidence of intensive participation on
articles and talk pages related to the events.  Furthermore, they
point out boosts of edits after major events, like Mubarak's
resignation. Another case study~\cite{Keegan2011} analyses edit and
collaboration dynamics around articles related to the 2011 Tohoku
earthquakes.

\begin{figure}[!tb]
  \centering
  \includegraphics[angle=-90,width=\columnwidth]{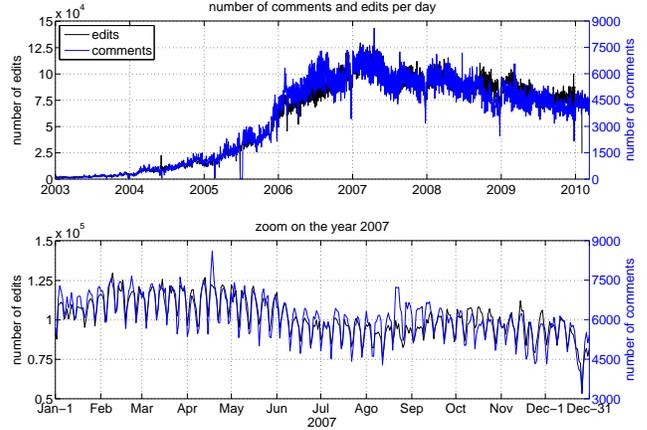}
  \caption{Evolution of the number of edits and comments per day
    between 2003 and 2010 (top) and a zoom on the activity during 2007
    (bottom subplot).}
  \label{fig:global_evolution}
\end{figure}

Temporal patterns in comment activity have also been investigated in
other contexts, e.g. for Slashdot comments
by~\cite{kaltenbrunner_LAWEB2007} who mainly analysed the reaction
time of a community to an initiator event. However, discussions in
Slashdot evolve on a much faster timescale than in Wikipedia. They
only receive a very short attention span and are subject to time
limits, making their dynamics highly influenced by the circadian cycle
of its authors. Activity peaks in social media platforms have been
studied in other contexts such as small text fragments from blogs and
mainstream media~\cite{leskovec2009meme},
Youtube~\cite{crane2008robust} and
Twitter~\cite{lehmann2011dynamical}. The two latter studies propose
models to characterise different classes of activity peaks according
to their dynamics, and to associate them to exogenous and endogenous
factors.

\section{Dataset and data preparation}
\label{sec:data}
We are interested not only in the evolution of the sizes of the
discussions but also in their complexity. To measure the complexity of
a discussion we will use its structure, thus we have to go beyond
simply analysing the edit history of article talk pages. We use a
dataset which identifies every comment and its associated metadata via
parsing the wikitext.

\begin{figure*}[!tb]
  \centering
  \includegraphics[angle=-90,width=\textwidth]{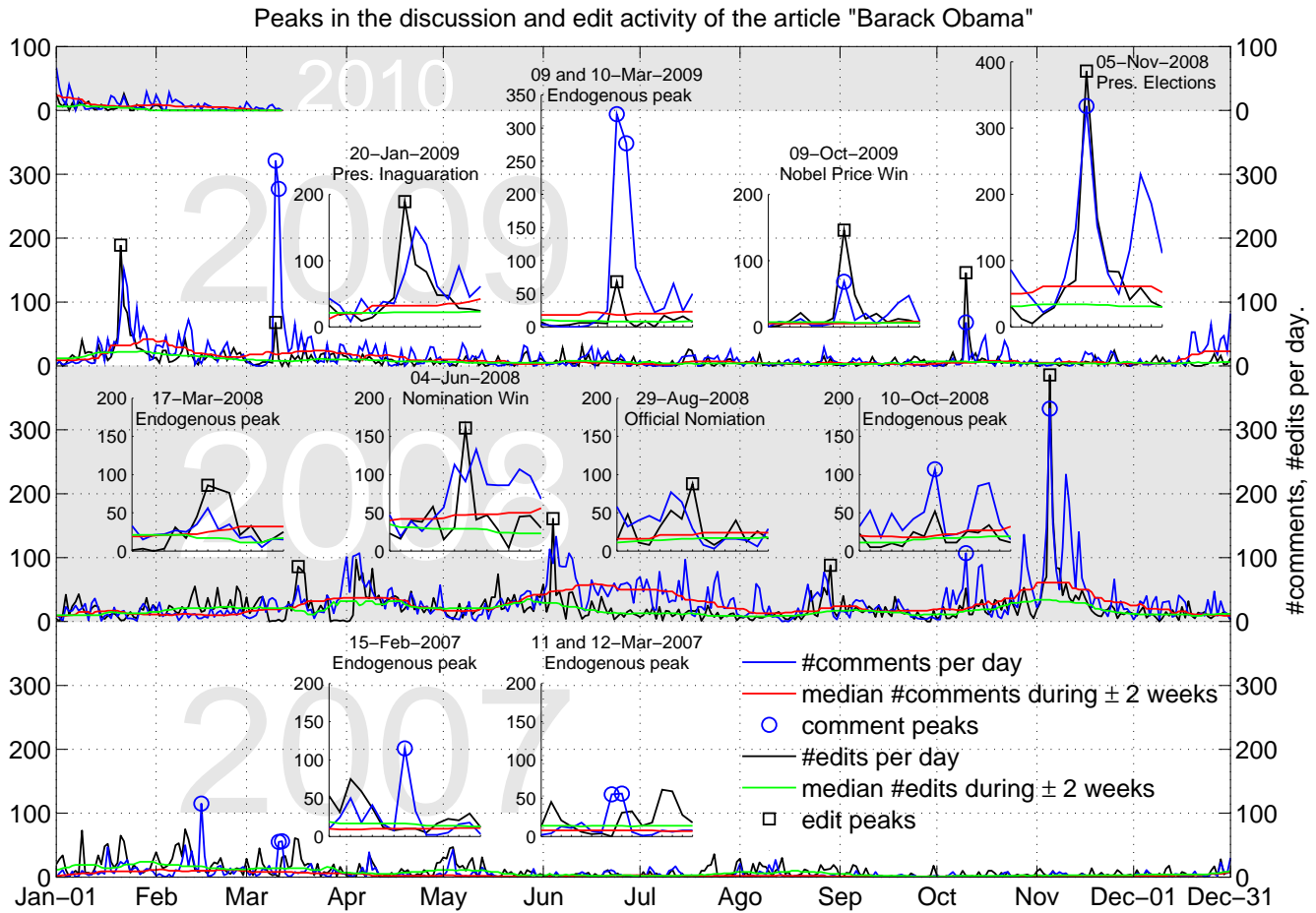}
  \caption{Peaks in the activity of the discussion around the article
    ``Barack Obama''.}
  \label{fig:President_peaks}
\end{figure*}

This dataset is described in more detail in~\cite{Laniado2011}.  It
contains all Wikipedia article talk pages, as of March 12th, 2010;
each discussion page is represented as a tree, to reflect the thread
structure of comments and replies indented under one
another\footnote{Note that sometimes users reset indentation; we
  consider these cases as the start of a new thread in the
  discussion.}.  Most comments are signed and dated by the
users\footnote{Or, since December 2006, by special bots automatically
  adding missing signatures and dates} adopting different conventions
and making use of wiki shortcuts. We only consider comments for which
an author-signature could be identified; of these 9.4 millions
comments, about 11\% have a malformed date-signature (or none at all).
We keep those comments as they still allow to extract the structure of
the discussion prior to the first correctly signed comment. This
information is then used as starting point for our statistics in the
second part of our study.

To extract the edit history of the articles we processed the complete
dump of Wikipedia including all revisions of each page, using the
WikiXRay
parser\footnote{\url{http://meta.wikimedia.org/wiki/WikiXRay}}.  We
only considered articles having a talk page ($\approx871,000$,
corresponding to 27\% of all articles in the
dump). Figure~\ref{fig:global_evolution} shows the evolution of the
number of edits and comments per day to these articles. We find about
6 comments for every 100 edits and a very similar trend in both
curves, which co-evolve nearly in perfect synchrony showing the same
weekly activity cycle.  The cycles becomes visible in the zoom on the
activity during 2007 in the bottom subplot of
Figure~\ref{fig:global_evolution}.\footnote{In this Figure the
  comments of two bots which made on rare occasions comments to
  several thousands different articles on a specific day were omitted
  for clarity.}

The first objective of this study is to investigate whether this
co-evolution between edits and comments can also be found at the
article level.

\section{Comment and edit peaks}
\label{sec:peaks}
We first analyse the co-occurrence of peaks in the comment and edit
activity of the different Wikipedia articles. We start by introducing
the method to detect the moments in time with peak activity.
\subsection{Peak detection algorithm}
We extract two time series for every Wikipedia article:
\begin{packed_enumerate}
\item The number of edits per day to the article.
\item The number of comments per day on the corresponding article talk
  page.
\end{packed_enumerate}
\vspace{-2mm} To identify peaks in these time-series we adapt the peak
detection algorithm of~\cite{lehmann2011dynamical}. We compare the
number $n(t)$ of edits (or comments) at day $t$ to the median activity
$m(t)$ during a sliding window of 4 weeks length\footnote{This length
  accounts for the weekly activity cycles~\cite{Yasseri2012} while
  reflecting seasonal (monthly) trends.} centred on day~$t$. We
consider the activity to peak if
\begin{equation}
  n(t) >c\cdot \max(m(t),n_{\min})
\label{eq:peak}
\end{equation}
where $n_{\min}$ is the minimum activity set to $10$ comments or $10$
edits respectively\footnote{The choice of 10 was taken in analogy
  to~\cite{lehmann2011dynamical}.} and $c$ is the peak-factor, which
we set to $5$ if not stated otherwise in the rest of the
manuscript. Consequences of other choices of $c$ are discussed in
Section~\ref{sec:parameters}.

We consider therefore the activity to peak at day $t$ if it is larger
than $c$ times the median activity during $\pm2$ weeks around $t$ (or
larger than $c$ times $n_{\min}$ if the median is lower than the limit
$n_{\min}$).

Note that this metric can also be seen as an extension of a method
presented in~\cite{ratkiewicz2010} where the logarithmic derivative,
i.e. the fraction $(n(t)-n(t-1))/n(t-1)$, was used to detect peaks in
the number of page views and incoming links of Wikipedia articles. Our
metric adds a sliding window and a minimum amount of activity to
obtain more stable results which do not depend on low activity
fluctuations (which are frequently found in the number of edits or
comments). Apart from that, a peak-condition of the logarithmic
derivative being larger than one, as used in~\cite{ratkiewicz2010},
would be equivalent to the use of $c=2$ in our case.

\subsection{A case study: the article ``Barack Obama''}
Before analysing the discussion and edit peaks on Wikipedia as a whole
we start first with a case study on a specific article to illustrate
our peak detection method as well as the co-occurrence of endogenous
and exogenous peaks.

Figure~\ref{fig:President_peaks} gives an example of the activity
peaks for the article ``Barack Obama'' between January 2007 and March
2010. The activity prior to 2007 is considerably lower and has been
omitted in the figure for clarity\footnote{Note that also February
  29th, 2008 has been excluded from the figure (but not from the
  analysis) to allow alignment with the non-leap years. No peaks were
  found for this day.}. The blue line corresponds to the number of
comments per day and the gray curve to the number of edits. The dashed
lines in red and green show the $\pm2$ weeks medians of the comments
and edits respectively. Peaks in the comment activity are depicted by
blue circles, edit peaks by black squares We count 6 comment peaks,
two of them correspond to twin peaks where the activity peaks on two
consecutive days. Our method also finds 6 (isolated) edit peaks, of
which 3 coincide in time with peaks in the comment activity. The other
3 edit peaks concur as well with elevated activity on the article talk
page but the number of comments involved does not surpass the
threshold established by Eq.~\ref{eq:peak}.

Some peaks clearly correspond to important events, such as Obama
winning the presidential primaries and elections, his inauguration and
the Nobel Peace Prize win. However, the subjects treated in the edit
peaks and intensive discussions are mostly not directly related to
these events, but focused on controversies regarding the figure of
Obama and the disputed neutrality of the article. This suggests that
these peaks of activity were mainly caused by attention peaks of the
public opinion towards the subject of the page.

\begin{figure}[!tb]
  \centering
  \includegraphics[angle=-90,width=\columnwidth]{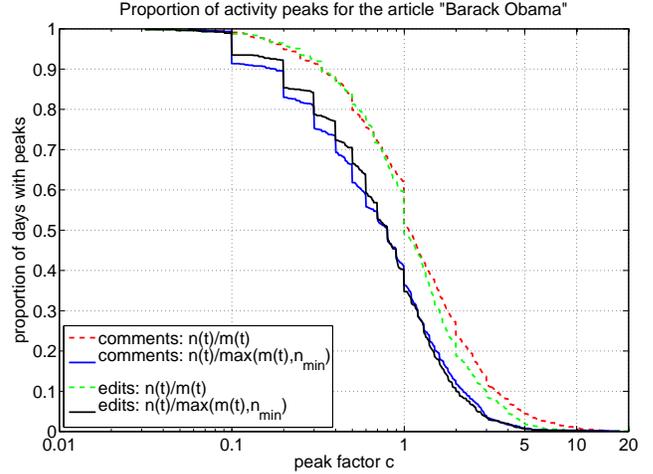}
  \caption{Proportion of days with peak activity for different choices
    of $c$ for the article ``Barack Obama''.}
  \label{fig:President_c}
\end{figure}

\begin{figure*}[!tb]
  \centering 

  \subfigure[Number of peaks per
  article]{\label{fig:peak_stats_A}\includegraphics[angle=-90,width=0.32\textwidth]{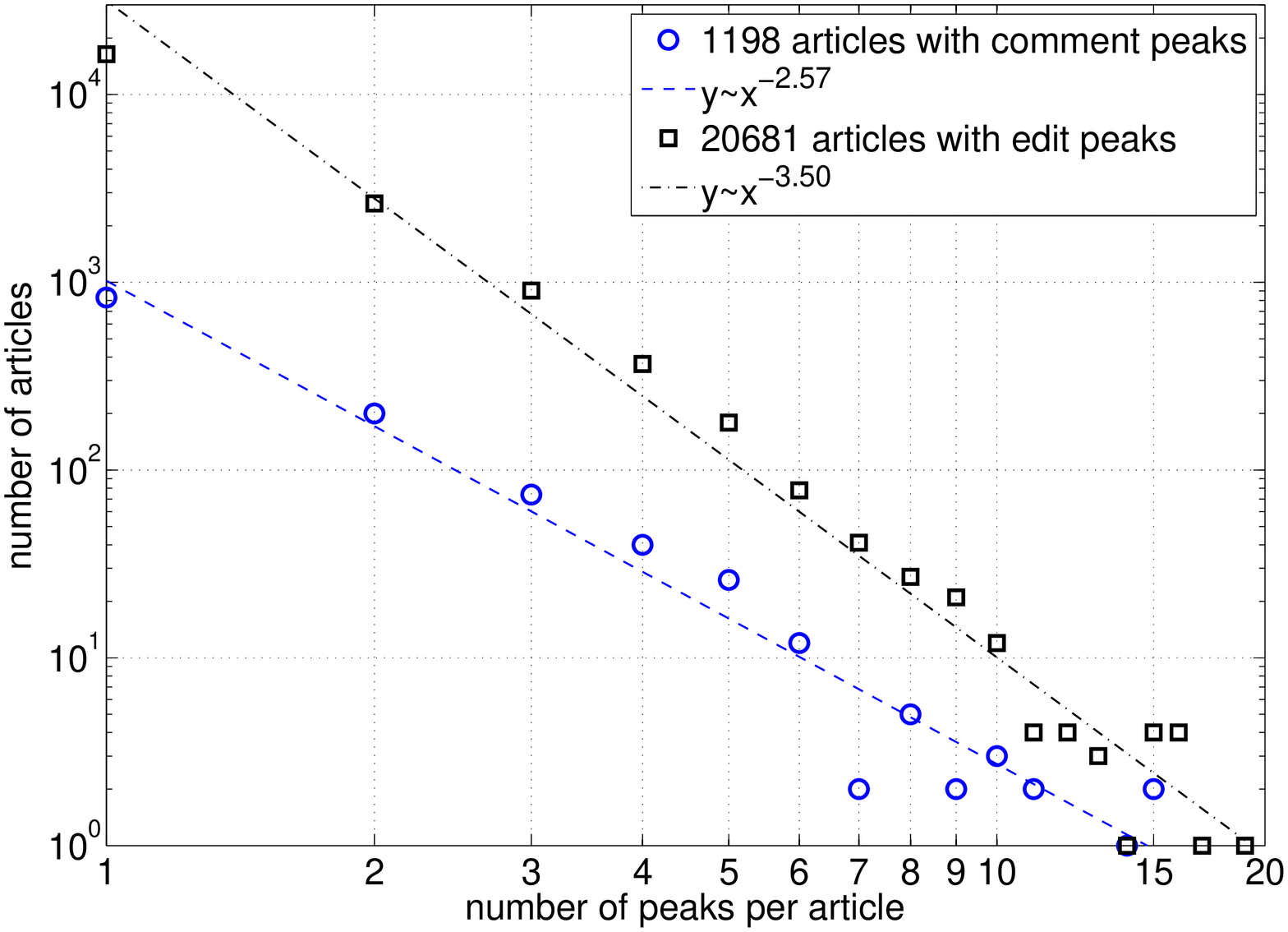}}
  \subfigure[peak-lengths]{\label{fig:peak_stats_B}
    \includegraphics[angle=-90,width=0.32\textwidth]{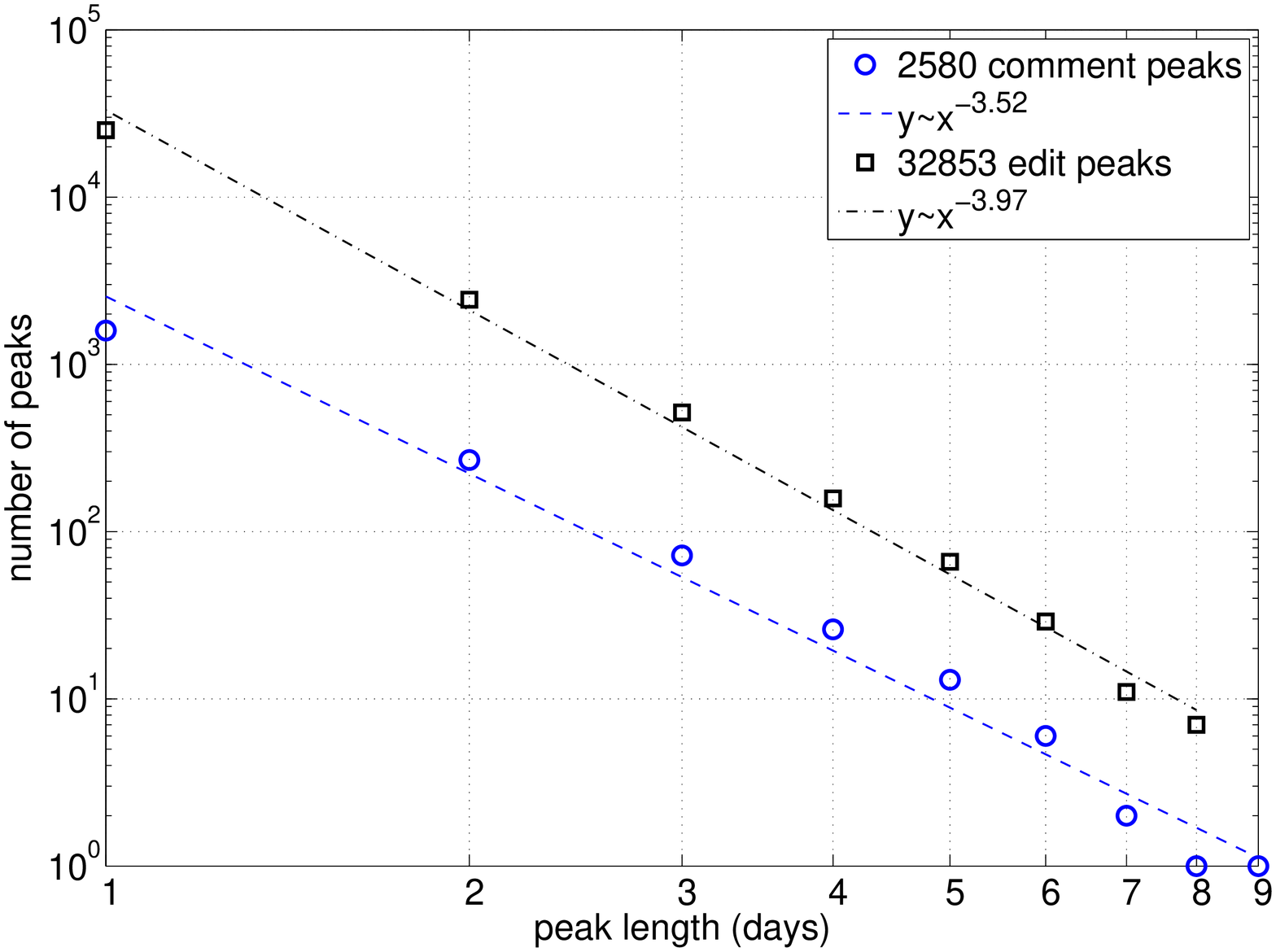}}
  \subfigure[time between two consecutive
  peaks]{\label{fig:peak_stats_C}
    \includegraphics[angle=-90,width=0.32\textwidth]{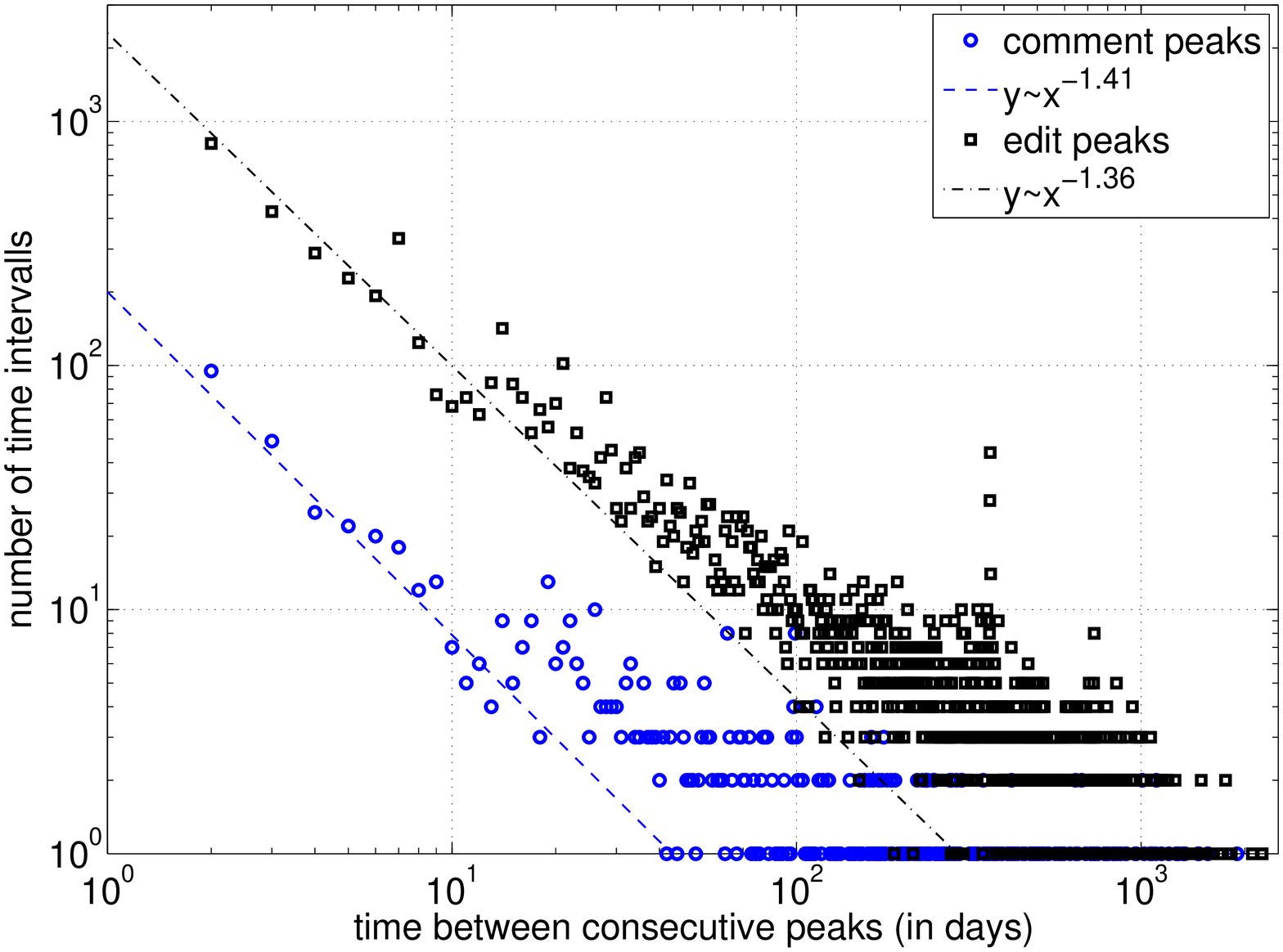}}
  \caption{Statistics of the edit (red squares) and comment peaks
    (blue circles).}
  \label{fig:peak_stats}
\end{figure*}

Other edit and comment peaks seem not to be related to major off-line
events, but caused by endogenous factors.  For example, the twin
discussion peak on March 11th and 12th, 2007 is due to a poll proposed
by a user to find consensus about which of the 13 main supposed
controversies, previously raised in the talk page, were
\emph{notable}\footnote{\url{http://en.wikipedia.org/wiki/Wikipedia:Notability}},
and thus had to be included in the article.  On March 9th, 2009, in
correspondence with a featured article
review\footnote{\url{http://en.wikipedia.org/wiki/Wikipedia:Featured_article_review}},
an edit war was raised by users trying to add to the article content
about controversial issues, such as Obama's contact with Rev. Jeremiah
Wright. The article was temporarily locked down, allowing only
administrators to edit, and rerouting the dispute to the talk
page. This explains the fall of edit activity after the peak on March
9th, accompanied by the burst of discussion in the talk page,
resulting in a twin peak (March 9th and 10th). The edit war was
reported by some conservative media like Fox
News\footnote{\url{http://www.foxnews.com/story/0,2933,507244,00.html}},
and media exposure amplified the expansion of the discussion.

Finally, we also observe two endogenous discussion peaks which occur
in the vicinity of external events but do not have a direct relation
with them. On February 15th, 2007 discussion peaks 5 days after
Obama's candidacy announcement and on October 10th, 2008 in between
the two presidential debates on the October 7th and 15th, 2008.

Overall, the example shows that both internal community dynamics and
external events can cause spikes of attention towards Wikipedia
articles, but the subjects of the disputes are not necessarily
directly related to the events which provoked them initially.

\subsection{Choice of the parameter values}
\label{sec:parameters}
As mentioned earlier we choose $c=5$, this way we ensure that peaks
are rare spiky events with a large difference from the usual activity
variations. For the article about ``Barack Obama'' the peaks we find
occur with a frequency of about~$1\%$. This can be observed in
Figure~\ref{fig:President_c} which shows the complementary cumulative
distribution of the ratio between the number of comments or edits
$n(t)$ and the 4 week medians $m(t)$ for all time-points (days) shown
in Figure~\ref{fig:President_peaks}. Continuous lines correct $m(t)$
with the minimum activity limit ($n_{\min}=10$) as in
Eq.~\ref{eq:peak} while dashed lines represent the pure ratio between
the two quantities\footnote{Which is more similar to the logarithmic
  derivative used in~\cite{ratkiewicz2010}.}. If the ratio surpasses
the peak-factor $c$ our method detects a peak. The curves roughly
resemble error functions in log-scale suggesting a log-normal
distribution of the variations around the median time series. We
observe that a choice of $c=2$ would lead to peaks in one of every 4
or 5 days without the use of $n_{\min}$, or one in 10 with correction.
Choices of $n_{\min}$ smaller than 10 would lead to intermediate
values. The choice of 10 was taken in analogy to
\cite{lehmann2011dynamical}.

\begin{table}[!tb]
\caption{Numbers of peaks and articles with peaks for different values of parameter $c$ in Eq.~\ref{eq:peak}.}
\label{tab:peak_stats}
\centering
\begin{tabular}{|l|r|r|r|} \hline
  peak-type & $c$  &num. of peaks &num. of articles \\
  \hline \hline
  comment  & 5 &2,580 &1,198\\
  & 10 &288 &195\\
  & 20 &30 & 21\\
  \hline
  edit & 5 & 32,853&20,681\\
  & 10 &4,944 &4,004\\
  & 20 &706 &631\\
  \hline
  \hline
  overlap & 5 & 307  & 59 \\
  same day& 10 & 44  & 8   \\
  & 20 &  7   & 3   \\
  \hline
  overlap & 5 & 703  & 385 \\
  $\pm 1$ day& 10 & 94  & 52   \\
  & 20 &  15   & 10   \\
  \hline
  overlap & 5 & 871  & 406 \\
  $\pm 2$ days& 10 & 121  & 54   \\
  & 20 & 19 & 11   \\
  \hline
\end{tabular}
\end{table}

Table~\ref{tab:peak_stats} lists the number of comment and edit-peaks
we obtained for our data set. We find 2,580 peaks in the discussions
and 32,853 edit peaks with $c=5$. Higher choices for $c$ significantly
reduce the number of peaks we detect: for $c=10$ we find roughly one
tenth of the comments peaks and 15\% of the edit-peaks. If we increase
$c$ further to $20$ the numbers decrease in similar proportion. One
can draw a similar figure as Figure~\ref{fig:President_c} (not shown)
to observe the effect of other choices of $c$. The proportions are
very similar when not correcting with the minimum activity limit
$n_{min}$, but are lower when using it. E.g. with $c=5$ we obtain, on
average, a peak for 0.2\% of the days with comment activity and for
0.5\% of the days with edit activity. This difference is caused by the
fact that the average activity per article is lower than the one for
the article ``Barack Obama''. Nevertheless, it is the articles with
high editing and commenting activity which have to be used to
calibrate the peak detection method to avoid an excessive number of
peaks.

\subsection{Peaks in the Wikipedia} 
In this section we present observations obtained when calculating peaks
in comment and edit activity for all English Wikipedia articles.

Once the peaks are detected the next natural question is whether this
edit and comment peaks coincide in time.  The bottom nine rows of
Table~\ref{tab:peak_stats} show that for $c=5$ only about $12\%$ of
all comment peaks coincide with an edit peak at the same day. This
number increases to $27\%$ when allowing one day of difference between
the edit and comment peaks and to $33.8\%$ when allowing two.  These
results indicate that not necessarily peaks in the discussion activity
have to lead to peaks in the editing activity as well. However, the
larger those peaks the more likely is a coincidence. E.g. for $c=20$
we find for $63\%$ of the comment peaks a corresponding edit peak
within at most 2 days distance.

Furthermore, we can observe that the average number of comment peaks
per article ($2.15$ for $c=5$ among the articles with at least one
peak) is larger than the corresponding number of edit peaks per
article ($1.59$ for $c=5$). An article with already one comment peak
is thus more likely to obtain a second one than an article with one
edit-peak.

\begin{table}[!tb]
\caption{Articles with at least two edit peak anniversaries. }
\label{tab:365peaks}
  \centering
  \begin{tabular}{|l|c|} \hline
 Title & \#edit-peak anniv.\\
\hline \hline
Boxing Day & 4 \\
Halloween & 3 \\
New Year's Eve & 3 \\
Guy Fawkes Night & 3 \\
May Day & 2 \\
Nowruz & 2 \\
Independence Day (United States) & 2 \\
Nickelodeon Kids' Choice Awards & 2 \\
\hline
\end{tabular}
\end{table}

This finding is confirmed when taking a deeper look at the
distribution of the number of peaks per article depicted in
Figure~\ref{fig:peak_stats_A}. We observe the typical shape of a
power-law-like distributions, with the majority of articles having at
most one peak. Although we find a considerably larger number of edit
peaks the shape of the distributions are similar for the comments
(blue circles) and edit peaks (black squares) with a steeper exponent
for the edit peak distribution. This implies (when normalising the
distributions) smaller likelihoods for having multiple edit peaks than
having multiple comment peaks.

We observe a similar characteristic for the distribution of the
peak-lengths, i.e. the number of consecutive days with peak activity,
in Figure~\ref{fig:peak_stats_B}, although in this case the difference
in the slopes is less pronounced.  The time between two consecutive
peaks also follows a power-law distribution as can be observed in
Figure~\ref{fig:peak_stats_C}.\footnote{ Note that we do not count
  peaks on consecutive days separately in this figure.} A maximum
likelihood estimation of the power-law exponents reveals exponents of
around $-1.4$. Further understanding of the shape of this inter-peak
time distributions in the Wikipedia community might be found in models
similar to the one of~\cite{barabasi05} which explain bursts and
inter-event time distributions of individual human behaviour.

\begin{table}[!tb]
\caption{Top 10 articles with most comment peaks.}
\label{tab:top10_numpeaks_comment}
  \centering
  \begin{tabular}{|l|c|c|} \hline
    Title & \#comment-peaks & \#edit-peaks  \\
    \hline \hline
    Intelligent design & 15  &2\\
    September 11 attacks &15  &3\\
    Race and intelligence &14 &5\\
    British Isles &11 &0\\
    Main page &11 &0\\
    Anarchism &10 &12\\
    Catholic church &10 &0\\
    Canada &10 &0\\
    Transnistria &9 &3\\
    New Anti-Semitism &9 &0\\
    \hline
\end{tabular}
\end{table}

\begin{table}[!tb]
\caption{Articles with the longest comment peaks.}
\label{tab:top10_maxpeakleangth_comment}
  \centering
  \begin{tabular}{|l|c|} \hline
    Title & max. comment-peak length \\
    \hline
    \hline
    2008-2009 Canadian  & 9 \\
    parliamentary dispute & \\
    \hline
    Seung-Hui Cho & 8 \\
    \hline
    Harry Potter and  &7 \\
    the Deathly Hallows & \\
    \hline
    Fort Hood shooting &7 \\
    \hline
    Capitalism &6 \\
    \hline
    Republic of Macedonia &6 \\
    \hline
    Bronze Soldier of Tallinn &6 \\
    \hline
    2008 South Ossetia war &6 \\
    \hline
    2008 Mumbai attacks &6 \\
    \hline
    July 2009 Ürümqi riots &6 \\
    \hline
\end{tabular}
\end{table}

\begin{figure*}[!tb]
  \centering \subfigure[number of comments per
  month]{\label{fig:President_comment_evolution_A}
    \includegraphics[angle=-90,width=0.49\textwidth]{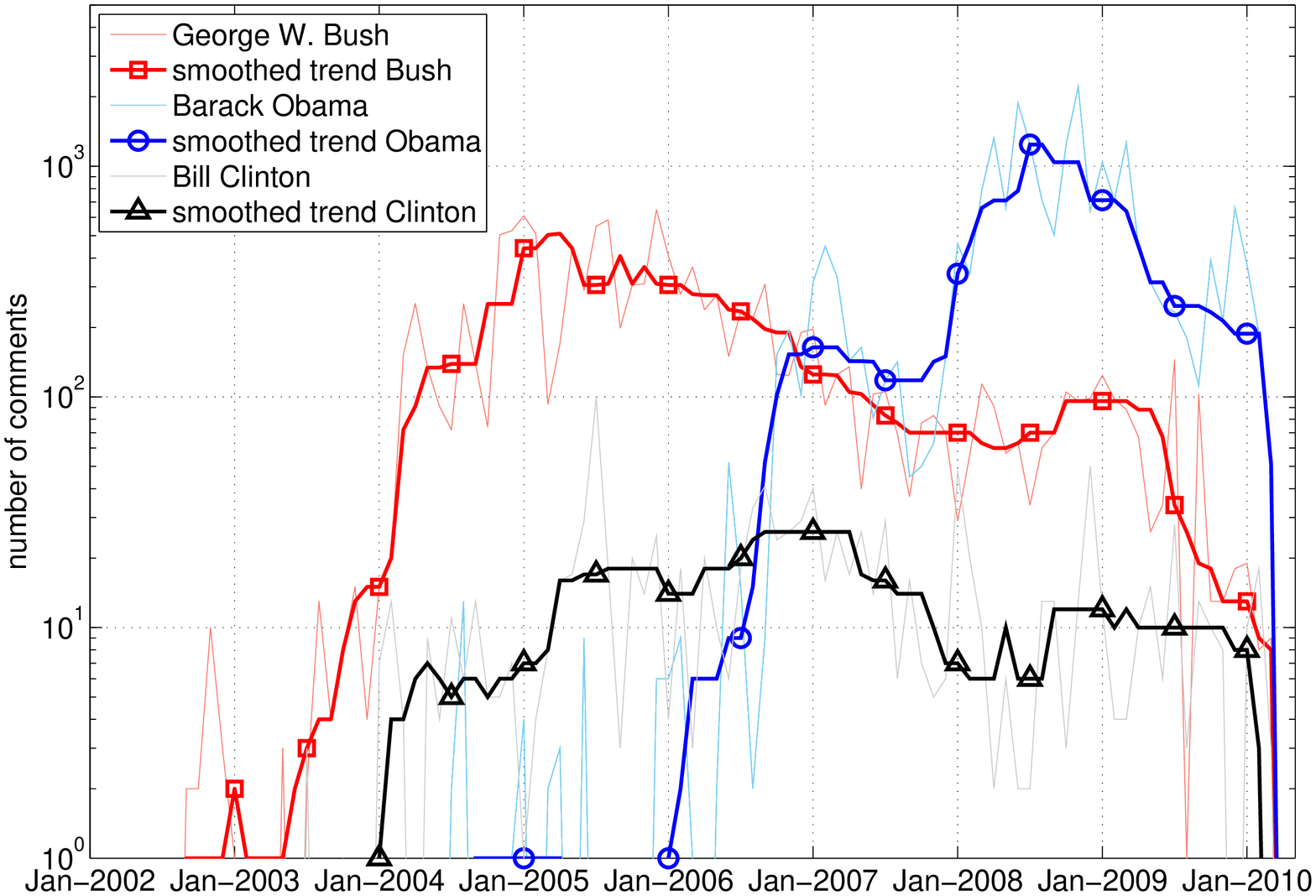}}
  \subfigure[growth of total number of
  comments]{\label{fig:President_comment_evolution_b}\includegraphics[angle=-90,width=0.49\textwidth]{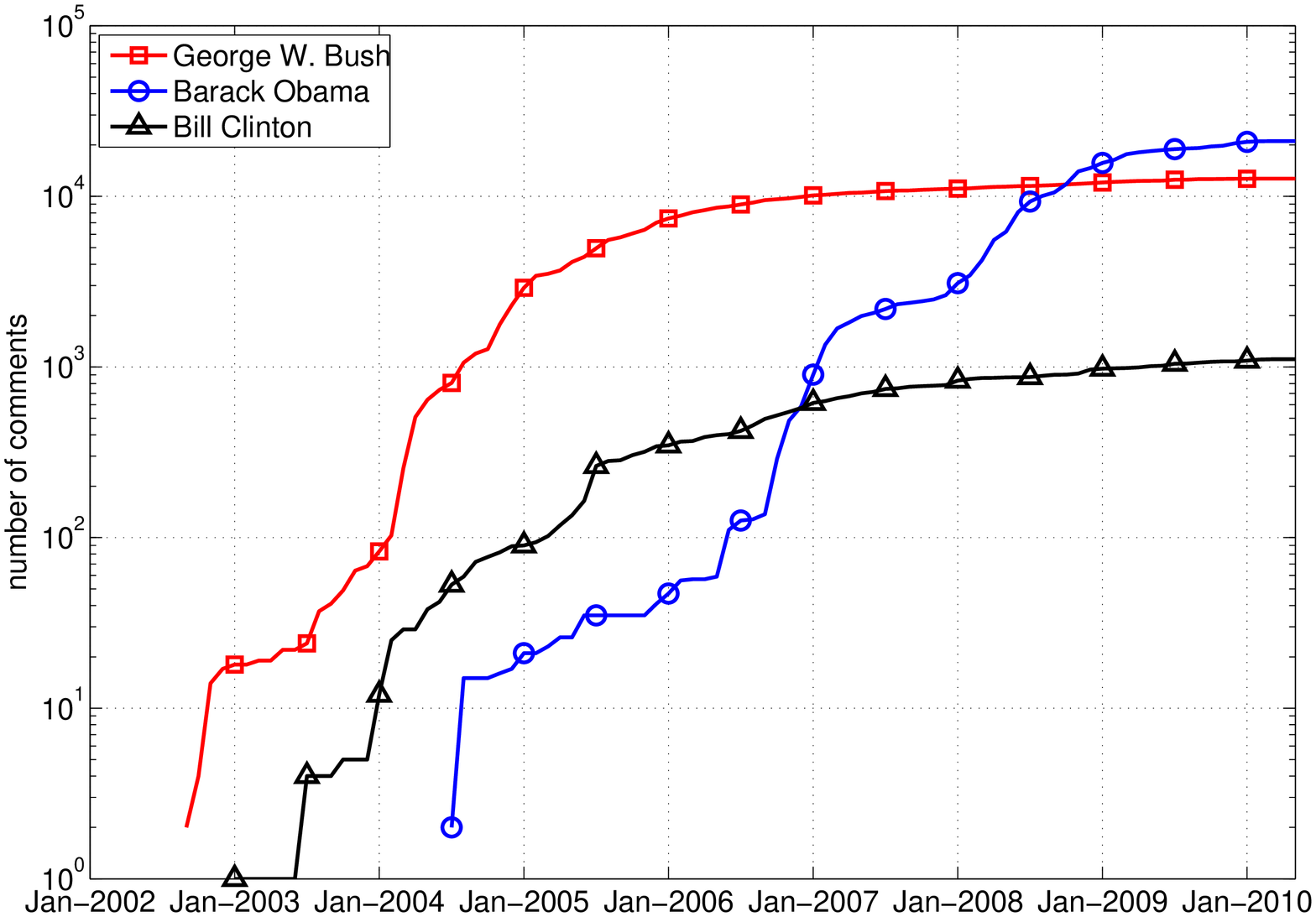}}
  \caption{The number of comments per month for the
    Wikipedia pages of the three most recent US-presidents.}
  \label{fig:President_comment_evolution}
\end{figure*}

Taking a closer look at Figure~\ref{fig:peak_stats_C} we furthermore
observe three outliers in the distribution of the edit peaks with
significantly more frequent peaks after 364, 365 and 366 days from the
previous one. This resonates with the anniversary effect described by
\cite{ferron2000studying} as people returning to articles at the
anniversary of a traumatic event. However, although we observe one
such repeated peak in the article about the September 11 attacks, most
of these ``anniversary peaks'' appear in articles about holidays or
other events with a fixed date in the calendar, as can be observed in
Table~\ref{tab:365peaks}. The table lists all articles that repeat an
edit peak at least two times. The anniversary effect is not visible in
the comment peaks and seems thus restricted to editing behaviour.

\begin{table}[!tb]
\caption{Top 10 articles with the most edit peaks.}
\label{tab:top10_numpeaks_edits}
  \centering
  \begin{tabular}{|l|c|c|} \hline
    Title & \#edit-peaks & \#com.-peaks  \\
    \hline \hline
    Uxbridge, Massachusetts & 19 & 0 \\
    \hline
    Voodoo (D'Angelo album) & 17 & 0 \\
    \hline
    List of World Wrestling  & 16 & 3 \\
    Entertainment employees  &  & \\
    \hline
    Super Smash Bros. Brawl & 16 & 2 \\
    \hline
    Michael Jackson & 16 & 1 \\
    \hline
    The Biggest Loser: & 16 & 0 \\
    Couples 2 & & \\
    \hline
    Roger Federer & 15 & 0 \\
    \hline
    Rafael Nadal & 15 & 0 \\
    \hline
    List of Barney \& Friends  & 15 & 0 \\
    episodes and videos & & \\
    \hline
    Total Drama Action & 15 & 0 \\
    \hline
\end{tabular}

\end{table}

We finish the section with three more tables. The first,
Table~\ref{tab:top10_numpeaks_comment}, shows a list of the top 10
articles with the largest number of the comment peaks and the second,
Table~\ref{tab:top10_maxpeakleangth_comment}, lists the 10 articles
with the longest peak-durations (consecutive days with peak-activity).
We find that many known controversial topics also have a large amount
of comment peaks; the top 7 articles listed in
Table~\ref{tab:top10_numpeaks_comment} can also be found among the
ones with the largest number of (prolonged) discussions between pairs
of users in a listing given in~\cite{Laniado2011}. However, with the
exception of the article about Anarchism, it is also interesting to
note that there can be found (if any) only a much lower number of edit
peaks in these articles. This might be explained by protection or
semi-protection of the corresponding articles. 

In the list of the longest comment peak durations given in
Table~\ref{tab:top10_maxpeakleangth_comment} we find articles (apart from
the one on Capitalism) about one time events, or with a clear relation
to an event such as the independence of a country or the publication
of a long awaited book. We will discuss some of these articles again
at the end of the next section.

Finally, we also list the top 10 articles with the largest number of
edit peaks in Table~\ref{tab:top10_numpeaks_edits}. Surprisingly, the
majority of the articles listed there seems to be only of limited
general interest, indicating that the corresponding edit peaks might
be caused by edit wars within a smaller group of users, or a single
very, very active user. The latter is for example the case in the
article about ``Uxbridge, Massachusetts''.

Tables~\ref{tab:top10_numpeaks_comment}
and~\ref{tab:top10_numpeaks_edits} suggest that the peaks in the
discussion activity are more suitable as a measure of a sudden, more
widespread interest of the Wikipedians in a specific topic than the edit
peaks. The latter seem to be easier to cause by a small group of people.
An adaptation of the measure taking into account the number of users
involved might be suitable to avoid this effect.

\subsection{Peak detection in real time}

So far we have presented a study of peaks in retrospective, as a first
necessary step for the understanding of edit and discussion
dynamics. However, the ability to detect peaks in real time could be
important for knowing what is going on in the wiki, and detecting
quick relative increments of activity around one or more articles.
This would be useful for the community, for example to involve
mediators early into a controversy and to try to prevent its
escalation into a too fierce debate.

To use our peak detection algorithm in real time, one would have to
modify the time window used to calculate the median activity, e.g. by
using just the activity during 2 weeks before the current
date. Another possible modification would be to monitor the fraction
of median and current activity instead of fixing a peak-factor. The
fraction could then be used to asses the relative size of a peak and
depending on the size different actions could be triggered.

\newpage
\section{The speed of  growth of the discussions}\label{sec:speed}
After having analysed peaks in the discussion activity we are now
interested in measuring how fast the discussions grow in complexity.

Figure~\ref{fig:President_comment_evolution} depicts the number of
comments received by the discussion pages of the Wikipedia articles of
the three most recent US-presidents. The left plot shows the evolution
of the number of comments per month, while the right plot corresponds
to the increase in the total number of comments on the pages. One can
observe that until the end of 2006 the page of George W. Bush received
more comments than those of his predecessor and successor. Between
January 2007 and October 2008 the page of G.W. Bush received a similar
amount of comments as the one of Barack Obama, who took the lead after
the elections of November 2008. The discussion page of the article
about Bill Clinton receives considerably less comments than those of
his two successors.

How should we assess now the complexity of the discussions on those
pages? For that purpose we will use the $h$-index of the structure of
the discussion introduced in~\cite{gomez2008slashdot}.

\subsection{The $h$-index of a discussion}
\begin{figure}[!tb]
  \centering
  \includegraphics[width=.65\columnwidth]{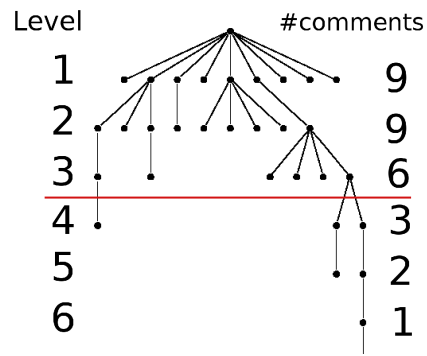}
  \caption{Example for the calculation of the $h$-index of a nested
    discussion with $h=3$, adapted from~\cite{gomez2008slashdot}.}
  \label{fig:h-index}
\end{figure}

The $h$-index of a discussion is a balanced depth
measure. Figure~\ref{fig:h-index} gives an example of the tree
representation of a nested discussion. The root node of the discussion
corresponds to its initiating event (i.e. the article in the case of
Wikipedia). The comments initiating a sub-thread %
are placed on the first level, their replies on the second and replies
to the replies on the third level and so forth. The level of the
comment is also refereed to as its depth.  The $h$-index of a
discussion is the maximal number $\theta$ such that there are at least
$\theta$ comments at level (depth) $\theta$, but not $\theta+1$
comments at level $\theta+1$. Another possible definition would be
that there are $\theta$ sub-threads of depth at least $\theta$.  The
red line in Figure~\ref{fig:h-index} indicates that for the example
thread the condition is fulfilled for $\theta=3$.

This measure is able to filter discussions which are very intense but
otherwise restricted to repeated arguments between two or a small
group of users.  Such threads can reach considerable depths but are
normally not representative to describe the complexity of the
discussions.  For example, the authors of~\cite{Laniado2011} found
discussions with a depth of 42 about the article ``Liberal
democracy'', while the $h$-index of this discussion was just $12$. The
maximal $h$-index they found was 20 for the discussion about
``Anarchism''.

\subsection{The growth measure $\Delta h$}
To measure how the $h$-index grows over time we define as $\Delta h$ the
average time (measured in days) it takes a discussion to increase its
$h$-index by one. Mathematically this can be written as: Given the
time-series $h(t)$ of the values of the $h$-index of a discussion
until day $t$ we search for the time-points $t_1,\ldots\,t_{\theta}$ such that
$h(t_i)=h(t_{t-1})+1$ for all $i \in [1,\ldots,{\theta}]$ and define then the
growth speed of the discussion as
\begin{equation}
\Delta h=\frac{\sum_{i=1}^{\theta-1}(t_{i+1}-t_{i})}{\theta-1}
=\frac{t_{\theta}-t_{1}}{{\theta-1}}.
\label{eq:deltah}
\end{equation}
The measure we propose here is very related to the inverse of the
$m$-index proposed in \cite{hirsch05}.  The $m$-index of a researcher
with and $h$-index of $\theta$ and who has first published a paper $n$
years ago is $m=\theta/n$. This definition takes advantage of the fact
that the $h$-index of a researcher should grow approximately linearly
in time. This is also true for the $h$-index of the discussions as we
will see in the next subsection.

\begin{figure}[!tb]
  \centering
\psfrag{xx2}[l][c][2][0]{George W. Bush $\Delta{h}=$70.7 days}
\psfrag{xx3}[l][c][2][0]{Barack Obama $\Delta{h}=$90.2 days}
\psfrag{xx4}[l][c][2][0]{Bill Clinton $\Delta{h}=$331.9 days}
\includegraphics[angle=-90,width=\columnwidth]{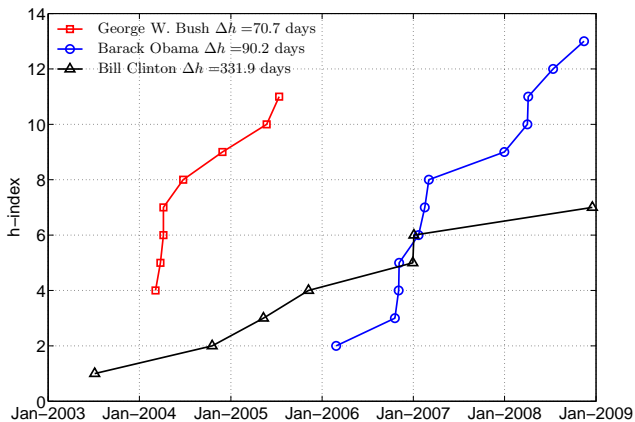}
  \caption{Evolution of the increase of the h-index for the three most
    recent US-presidents.}
  \label{fig:President_h_evolution}
\end{figure}

\subsection{Results for three examples}
Once introduced we can calculate the growth measure for the discussion
pages of the three most recent US-presidents.
Figure~\ref{fig:President_h_evolution} shows the increase in time of
the $h$-index of these three pages. We observe a more or less constant
growth of the discussions, validating the linearity assumed in the
definition of Eq.~\ref{eq:deltah}. Note that as explained in the data
section the date of the older comments in Wikipedia could not always
be determined due to format issues. This explains why the curves in
Figure~\ref{fig:President_h_evolution} do not start for all articles
with $h=1$ (in that case Eq.~\ref{eq:deltah} has to be adapted to
average over less time-intervals). For example, for the article about
George W. Bush we can only determine the time-stamp of the comments
when the $h$-index of the discussion already reached an $h$-index of
4. We observe that the articles about the two presidents in office
during the time since Wikipedia has existed experience a considerable
faster growth than the page about their predecessor Bill Clinton. For
George W. Bush we observe an average $\Delta{h}$ of 70.7 days, for
Barack Obama\footnote{Note here that in difference to
  \cite{Laniado2011} we ignore here structural elements such as
  headlines in the calculation of the $h$-index which explains the
  smaller final $h$-index compared to the ones reported
  by~\cite{Laniado2011}.} this value is 90.2 days, while the article
about Bill Clinton takes on average 331.9 days to increase its
$h$-index by one. This might again be explained by a considerably
larger influence of recent events on the discussion dynamics.

\subsection{General results}

How do these growth rates compare with those of other articles? To
answer this, in this section we take a look at the distribution of
$\Delta{h}$ for all $826$ articles with more than 1000 comments in our
dataset.

\input{table_bottom20_WikiSym}
\input{table_top20_WikiSym}

\begin{figure}[!tb]
  \centering
\psfrag{xx1}[c][c][2][0]{$\Delta{h}$}
  \includegraphics[angle=-90,width=\columnwidth]{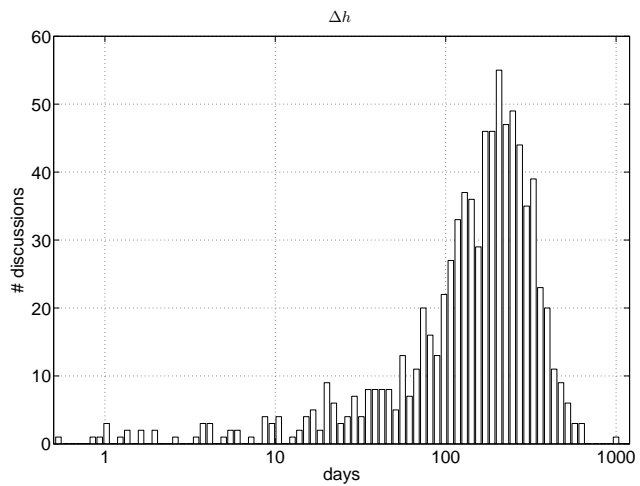}
  \caption{Distribution (logarithmic binning) of the $\Delta{h}$ of
    all discussions with more than 1000 comments.}
  \label{fig:delta_h_distribution}
\end{figure}

Figure~\ref{fig:delta_h_distribution} depicts this distribution (with
logarithmic binning). We observe a certain resemblance to a Gaussian
shape, with the mode of the distribution at values of $\Delta{h}$ of
around 200 days (mean =183.1 and median =172.7).  The articles about
Barack Obama and George W. Bush are in the quartile of the fastest
growing discussions, while the one about Bill Clinton can be found in
the decile of the slowest discussions.  However, none of these
discussions falls into the extremes of very slow or very fast growing
discussions.

Tables \ref{tab:sort_fastest20} and \ref{tab:sort_slowest20} list the
top discussions according to these criteria together with their
$\Delta h$ values. The start date indicates the earliest dated
comment\footnote{As mentioned above this does not have to mean that
  the discussion started that day, as about 11\% of the comments, and
  especially the oldest comments, do not have a date associated in our
  dataset.}, the end date the day when the the final $h$-index has
been reached for the first time (note that this will be in most cases
not the day of the last comment).  We observe that many of the fastest
evolving discussions appear around articles related to events which
received heavy news coverage, such as school shootings (the Virginia
Tech massacre and its author which occupy ranks~1 and 5 in
Table~\ref{tab:sort_fastest20}), the 2009 flu pandemic, terrorist
attacks, air crashes, etc. Nevertheless we find also topics which
reflect ideological or ethical motivated disputes among the Wikipedia
editors which lead to discussion gaining complexity very fast. Such
topics are the ``Bronze Solder of Tallinn'' (reflecting a conflict
between ethnic Russians and Estonians), ``the 2009 Honduran
constitutional crisis'' as well as discussions about the ``Israeli
occupied territories'' and the ``International status of Abkhazia and
South Ossetia''. Some of the articles of this list have already
appeared in Table~\ref{tab:top10_maxpeakleangth_comment} suggesting a
correlation between $\Delta{h}$ and the maximum edit peak-length of
the articles. Such a correlation can indeed be found ($r=-0.25$, weak
although statistically significant, $p<10^{-4}$) indicating that
faster growing discussions are more likely to have longer lasting edit
peaks.

We also find the very similar topics ``State Terrorism by the United
States'' and ``State terrorism and the United States'' appearing in
the list. They correspond in fact to the same article, which has been
renamed several times (the current title is ``United States and state
terrorism'') leaving only the archived discussions under the old
titles. We have decided to keep these separated discussions, to show
the repeated fast growth of the discussion around the slightly renamed
(and re-framed) subject.

Finally the list of the slowest evolving discussions in
Table~\ref{tab:sort_fastest20} is led by the articles about
``Christopher Columbus'' and ``Pi'' and contains many more articles
about timeless content or content which has been subject of
discussion over prolonged time such as ``Harry Potter'' or the ``War on
Terrorism''.  Some of these topics may well be topics of century long 
dispute such as the ``Shakespeare authorship question''.

\subsection{Using $\Delta h$ to measure  discussion maturity}

Our results show that article discussions can have very different
dynamics, with time-spans ranging from days to years.  Also,
Figure~\ref{fig:President_h_evolution} suggests a remarkable
difference between discussions which are still growing in complexity,
as the one about the article ``Barack Obama`'', and those whose
$h$-index has not increased recently, like in the case of the article
``George W. Bush''. It seems reasonable to assume that, although both
articles are still being edited and discussed, the content of the
second one should be more stable as its discussion is mature in the
sense that its most conflictive controversies have been fought,
i.e. its maximal $h$-index has been reached.  That does not mean that
no new controversy might emerge somewhere in the future to raise it
further.

The growth measure $\Delta h$ introduced in this section can be used
to assess this \emph{maturity} of a discussion.  One has to assure
that a sufficient amount of time has passed since the last time the
$h$-index of a discussion has increased. Only after some multiples of
$\Delta h$ one can safely assume that the discussion will not grow
more in complexity unless something unexpected will happen.  However,
more research is need to investigate the correct parametrisation of
the proposed technique for this specific application.

\section{Conclusions}
In this study we have analysed the time-evolution of the discussions
about Wikipedia articles on two scales: at a microscopic level
detecting peaks in the comment activity and at a macroscopic level
introducing a measure for the rate of increase in complexity of the
discussions. Although other studies have systematically analysed peaks
on Wikipedia before~\cite{ratkiewicz2010}, to our knowledge this is
the first study which does this for the commenting and editing
activity.

We observe that, although on a global scale editing and commenting
activity co-evolve nearly in perfect synchrony, at the individual
article level discussion and edit peaks seem to occur mostly
independently of each other. They can be caused by either endogenous
dynamics, such as polls, internal peer reviews and initiatives of some
editors, or exogenous factors, such as off-line events or media
exposure.  However, the subjects of the disputes are not necessarily
related directly to the events which provoked a peak of attention.

By introducing a measure for the growth of a discussion based on a
balanced depth-measure of its structure we find several orders of
magnitude of different rates of increase in complexity of the
discussions. Some discussions need only a few days to evolve, while
the slowest go on over years.

Our findings indicate that current events are often edited and
discussed on Wikipedia about nearly in real-time either in a new
article or in the form of a new peak in an already existing article,
while on the other hand, for articles about historical or scientific
facts which are not on the people's minds it may take years to reach a
similar level of intensity.

Wikipedia is the most up-to-date encyclopedia seen until now and is
heavily influenced by recent events as can be observed for example
when consulting the
Wiki-Zeitgeist\footnote{\url{http://stats.wikimedia.org/EN/TablesWikipediaEN.htm#zeitgeist}},
i.e. the list of the most edited topics in a certain time period.
Using the metrics presented here one should be able to develop more
sophisticated algorithms for such rankings which instead of just
counting edits would account also for relative growth dynamics in both
edit and discussion activity.

Although there is concern inside the Wikipedia community about the
risks of \emph{recentism} and the disproportionate attention received
by recent and current events, our results point out that this type of
dynamics constitutes a characteristic pattern of activity in the
Wikipedia community. Such patterns are probably unavoidable in the
context of an open collaborative platform.

According to the interpretation of Wikipedia as a collective memory
place~\cite{pentzold2009fixing}, where a \emph{living memory} is
progressively turned into a fixed text, talk pages are the space where
this transformation takes place.  Wikipedia discussions can thus be
seen as a mirror of a stream of public consciousness, where those
elements which are still not part of a shared consolidated heritage
are object of a continuous negotiation among different points of view.

\mpara{Future Work} Our study presents a first step towards the
comprehension of the temporal patterns which rule the collective
process of content creation in Wikipedia, and opens up to further
research in several directions.

We have sketched applications of our metrics for early detection
of controversies and assessment of discussion maturity, which should
be implemented for further evaluation.

The extension of existing generative models~\cite{gomez2012likelihood}
for the tree structure of the discussions with temporal information
could help to explain possible dynamical differences between fast and
slow evolving discussions.

A cross-article analysis of peaks would help to unveil hidden
relationships between articles and topics, while the classification of
peaks according to their shape, as for example proposed
in~\cite{crane2008robust}\footnote{This would make necessary a more
  fine grained temporal analysis with time rescaling to avoid the
  influence of daily activity cycles.}, could allow to automatically
distinguish exogenous and endogenous peaks, and to detect the
influence of external events.  The metrics used in this study could
also be applied on specific sub-threads of a discussion, which should
allow to detect hot spots within a specific article.

\section*{Acknowledgements} 

This work was partially supported by the Spanish Centre for the
Development of Industrial Technology under the CENIT program, project
CEN-20101037, “Social Media”.

 \bibliographystyle{abbrv}

\balancecolumns
\end{document}

%% file: table_bottom20_WikiSym.tex
\begin{table*}[!htb]
\caption{The 15 fastest discussions, $\Delta h$ and duration are given in days.}
\label{tab:sort_fastest20}
  \centering
  \begin{tabular}{|p{7.3cm}|c|r|r|c|c|} \hline
 Title & ~~$\Delta{h}$~~ & start date~~ & end date~~ & duration & final h-index \\
\hline \hline
Virginia Tech massacre & 0.5 & 15-Apr-2007 & 20-Apr-2007 & 5 & 9 \\
2009 flu pandemic & 0.9 & 25-Apr-2009 & 30-Apr-2009 & 5 & 7 \\
Bronze Soldier of Tallinn & 0.9 & 26-Apr-2007 & 02-May-2007 & 6 & 7 \\
2009 Honduran constitutional crisis & 1.0 & 27-Jun-2009 & 05-Jul-2009 & 8 & 8 \\
Seung-Hui Cho & 1.0 & 16-Apr-2007 & 24-Apr-2007 & 8 & 8 \\
2008 Mumbai attacks & 1.0 & 26-Nov-2008 & 01-Dec-2008 & 5 & 6 \\
Israeli-occupied territories & 1.2 & 22-Sep-2005 & 03-Oct-2005 & 11 & 10 \\
International status of Abkhazia and South Ossetia & 1.3 & 25-Aug-2008 & 04-Sep-2008 & 10 & 8 \\
Air France Flight 447 & 1.4 & 01-Jun-2009 & 08-Jun-2009 & 7 & 6 \\
7 July 2005 London bombings & 1.7 & 10-Jul-2005 & 15-Jul-2005 & 5 & 5 \\
State terrorism and the United States & 1.7 & 15-Feb-2008 & 06-Mar-2008 & 20 & 13 \\
July 2009 Ürümqi riots & 1.9 & 06-Jul-2009 & 21-Jul-2009 & 15 & 9 \\
Henry Louis Gates arrest controversy & 2.0 & 24-Jul-2009 & 09-Aug-2009 & 16 & 9 \\
Teach the Controversy & 2.6 & 11-Apr-2005 & 29-Apr-2005 & 18 & 8 \\
State Terrorism by the United States & 3.3 & 31-May-2007 & 03-Jul-2007 & 33 & 11 \\
\hline
\end{tabular}
\end{table*}

%% file: table_top20_WikiSym.tex
\begin{table*}[!htb]
\caption{The 15 slowest discussions, $\Delta h$ and duration are given in days.}
\label{tab:sort_slowest20}
  \centering
  \begin{tabular}{|p{7.3cm}|c|r|r|c|c|} \hline
 Title & ~~$\Delta{h}$~~  & start date~~ & end date~~ & duration & final h-index \\
\hline \hline
Christopher Columbus & 1159.0 & 24-Oct-2003 & 27-Feb-2010 & 2318 & 5 \\
Pi & 627.3 & 07-Dec-2002 & 20-Oct-2009 & 2509 & 6 \\
New York City & 617.3 & 09-Dec-2003 & 03-Jan-2009 & 1852 & 5 \\
Anna Anderson & 604.5 & 17-Mar-2004 & 09-Jul-2007 & 1209 & 3 \\
Harry Potter & 589.9 & 27-Nov-2002 & 02-Oct-2007 & 1770 & 6 \\
France & 566.5 & 13-Nov-2003 & 26-Jan-2010 & 2266 & 6 \\
Scientific method & 553.5 & 15-Jun-2003 & 08-Jul-2009 & 2215 & 6 \\
Instant-runoff voting & 546.3 & 09-Jul-2003 & 03-Jan-2008 & 1639 & 5 \\
Fathers' rights movement & 546.0 & 07-Mar-2004 & 01-Sep-2008 & 1639 & 4 \\
War on Terrorism & 533.1 & 07-Oct-2005 & 22-Feb-2010 & 1599 & 6 \\
World War II casualties & 523.0 & 13-Sep-2004 & 29-Dec-2008 & 1568 & 4 \\
Vampire & 517.1 & 19-Nov-2002 & 18-Jul-2008 & 2068 & 6 \\
Led Zeppelin & 511.9 & 31-Jan-2003 & 03-Feb-2010 & 2560 & 6 \\
Karl Marx & 487.6 & 19-Sep-2004 & 21-Jan-2010 & 1950 & 6 \\
Shakespeare authorship question & 485.6 & 02-Jun-2003 & 24-Jan-2010 & 2428 & 7 \\
\hline
\end{tabular}
\end{table*}